\documentclass[12pt]{article}

\usepackage{amssymb}
 \usepackage{amsthm}
\usepackage{multirow}
\usepackage[normalem]{ulem}
\usepackage{url}

\usepackage[pdftex]{graphicx}
\usepackage[tight,footnotesize]{subfigure}
\graphicspath{{figures/}}
  \DeclareGraphicsExtensions{.pdf,.jpg,.png}

\begin{document}


\title{New Thread Migration Strategies for NUMA Systems}

\author{
	O. G. Lorenzo\textsuperscript{1}, M. L. Beco\~{n}a\textsuperscript{1}, T. F. Pena\textsuperscript{1},\\ J. C. Cabaleiro\textsuperscript{1}, J. A. Lorenzo\textsuperscript{2} and F. F. Rivera\textsuperscript{1}\\
\footnotesize \textsc{\textsuperscript{1}CiTIUS Centro de Investigaci\'{o}n en Tecnolox\'{i}as da Informaci\'{o}n}\\
\footnotesize \textsc{Universidade de Santiago de Compostela, Spain}\\
\footnotesize \textsc{\textsuperscript{2}Quartz Research Lab - EISTI Pau, France}
}

\maketitle

\begin{abstract}
Multicore systems present on-board memory hierarchies and communication networks that influence performance when executing shared memory parallel codes.
Characterising this influence is complex, and understanding the effect of particular hardware configurations on different codes is of paramount importance.
In previous works, monitoring information extracted from hardware counters at runtime has been used to characterise the behaviour of each thread in the parallel code in terms of the number of floating point operations per second, operational intensity, and latency of memory access.
We propose to use this information to guide thread migration strategies that improve execution efficiency by increasing locality and affinity.
Different configurations of NAS Parallel OpenMP benchmarks on multicores were used to validate the benefits of the proposed thread migration strategies.
Our proposed strategies produce up to 70\% improvement in scenarios where locality and affinity are low, there being a small degradation in performance for codes with high locality and affinity.

\end{abstract}


\section{Introduction}
Current microprocessors implement multicores that feature a diverse set of compute cores and on board memory hierarchies connected by increasingly complex communication networks and protocols with area, energy, and performance implications.
For a parallel code to be correctly and efficiently executed in a multicore system, it must be carefully programmed, and memory sharing stands out as a sine qua non for general purpose programming~\cite{sodan2005}. A critical programming challenge for these systems is to partition application tasks, mapping them to one of many possible core thread configurations to achieve a desired performance in terms of throughput, delay, power, and resource consumption, among others~\cite{Ju2014}. The number of mapping choices increases as the number of cores and threads increase.

Considering the architectural features, particularly those that determine the behaviour of memory access, it is critical to improve locality of access and affinity among threads, data, and cores.
Performance issues that are impacted by this information are, among others, data locality, thread affinity, and load balancing, and so addressing these issues is critical to improve performance in general~\cite{PDPTA-MemoryAccessHC}.

Various performance models have been proposed to understand the performance of a code running on a particular system~\cite{moore2001review,  adhianto2010hpctoolkit, morris2008observing, geimer2010scalasca, cheung2008performance, mohr2002performance, schulz2007pn}.
In particular, the roofline model (RM)~\cite{Williams:2009:RIV:1498765.1498785} offers a balance between simplicity and descriptiveness based on the operational intensity (OI), defined as the number of operations per byte of DRAM traffic, measured in floating point operations per second (FLOPS)/Byte (flopsB); and the number of FLOPS, measured in GFLOPS.
The original RM presented drawbacks that have been previously analysed~\cite{lorenzo2013,lorenzo2014,lorenzoGranada2014}.
This, the dynamic roofline model (DyRM) was proposed~\cite{lorenzo2013}, essentially the equivalent of splitting the execution of a code in time slices, getting one RM for each slice, and then combining them in a single graph that shows the evolution of the code when running.
The latency extended DyRM (3DyRM)~\cite{lorenzo2014, lorenzoGranada2014} extended the  DyRM model with an additional parameter, memory access latency, measured in number of cycles.
These works also detailed how the information provided by Precise Event Based Sampling (PEBS)~\cite{Intel2014,pebs} on Intel processors was processed to obtain parameters that defined the models (flopsB, GFLOPS, and latency).
Even though these parameters are related, they incorporate the important factors that influence performance of parallel shared memory code when executed in a shared memory system, and in particular, in multicores.

Moving threads close to where their data reside can help alleviate memory related performance issues, since when threads migrate, the corresponding data usually stays in the original memory module, and is accessed remotely by the migrated thread~\cite{Chasparis2017}.
This could induce inefficiency that, sometimes, cannot be alleviated by the benefits of the migration~\cite{constantinou2005, shim2012judicious, sibai2010simulation, li2013numa, klug2011autopin}. 
Some analytical results are available for multicore processor analysis. For example,~\cite{Guz2009} performed a mean value analysis of a multithreaded multicore processor and showed that there is a performance valley to be avoided as the number of threads increases. Markovian models were used in~\cite{Chen2009} to model a cache memory subsystem with multithreading, and other works~\cite{Bhaskar2008, Ju2014} have modelled multithreaded multicore using queuing models.

We use the 3DyRM model to implement strategies for migrating threads in shared memory systems and, in particular, multicores.
The concept is to use the defining parameters of 3DyRM as objective functions to be optimised.
Thus, it is considered as a multiobjective optimisation problem.
The proposed technique is an iterative method inspired from evolutionary optimisation algorithms.
To this end, we define an individual utility function to represent the relative importance of the 3DyRM parameters.
This function is a weighted product that can be considered as representative of the performance of each parallel thread, and the parameters characterise the efficiency of each thread. Thus, a single value is able to quantify the performance of each thread in terms of locality and affinity.

Section~\ref{sec:parameters} describes the 3DyRM parameters used to characterise the execution of each thread in the parallel code.
We also summarise the use of hardware counters to extract the information required for the 3DyRM with low overhead.
 Section~\ref{sec:migr} introduces the proposed thread migration strategies. 
 A set of case studies based on the NAS Parallel OpenMP benchmarks (NPB-OMP)~\cite{bailey1991parallel} are described in Section~\ref{sec:CasesStudy}, and  the outcomes are discussed. Finally, the main conclusions are summarised in Section~\ref{sec:Conclusions}. 

\section{Parameters to characterise the performance of threads in parallel code}
\label{sec:parameters}
In modern systems the main bottleneck is often the connection between the processor(s) and memory~\cite{mckee2004reflections}.
3DyRM relates processor performance to off-chip memory traffic.
Operational intensity (OI) is the operations per byte of DRAM traffic (measured in flopsB). OI measures traffic between the caches and main memory rather than between the processor and caches.
OI incorporates the DRAM bandwidth required by a processor in a particular computer, and the cache hierarchy, since better use of cache memories would mean less use of main memory.
Thus, DyRM brings together floating point performance, operational intensity, and memory bandwidth.
However,  OI is insufficient to fully characterise memory performance, particularly on non-uniform memory access (NUMA) systems. In a NUMA system, distance and connection to memory cells from different cores may induce variations in memory latency, and so the same code may perform differently depending on where it was scheduled, which may not be detectable in DyRM.
Extending DyRM with the mean latency of memory access provides a better model of performance.
Thus, we employ the 3DyRM model, which provides a three dimensional representation of thread performance on a particular placement.

PEBS is an advanced sampling feature of Intel Core based processors, where the processor directly recording samples from specific hardware counters into a designated memory region.
The use of PEBS as a tool to monitor a program execution and perform thread migrations was implemented by \cite{lorenzo2014multiobjective}, providing  runtime dynamic information about code behaviour with low overhead~\cite{Akiyama:2017:QEI:3095770.3095773,Intel2014}.
Our migration tool constantly gathers performance data in terms of the 3DyRM parameters, GFLOPS, flopsB, and latency, for each core and thread.

However, the floating point (FP) information from PEBS may sometimes be inaccurate. FP instructions may be counted more than once when the memory is used intensively, because they are counted when issued, not when retired, and if their operands are not available on the L1 cache they may be issued more than once until they are read from higher memory levels~\cite{Intel2014}.
Additionally, new models of Intel processors do not allow direct reading of FP operations.
Therefore, information about retired instructions is also recorded, so giga instructions per second (GIPS) and instructions retired per byte (instB) may be used rather than GFLOPS and flopsB, respectively, in relevant cases.

\section{A new thread migration strategy}
\label{sec:migr}
We introduce a new strategy for guiding thread migration in NUMA systems.
The proposed algorithm performs threads migrations iteratively each $T$ milliseconds.
The concept is to consider the 3DyRM parameters as objective functions to be optimised, so that that increasing GFLOPS (or GIPS) and flopsB (or instB), and decreasing latency in each thread improves performance in the parallel code.
There is a close relation between this and multiobjective optimisation (MOO) problems, which have been extensively studied~\cite{SurveyMOO2017}. The aim of most MOO solutions is to obtain the Pareto optimality numerically. However, this task is usually computationally intensive, and consequently a number of heuristic approaches have been proposed. 

In our case, there are no functions to be optimised. Rather, we have a set of values that are continuously measured in the system.
Therefore, we propose to apply MOO methods to address the problem using the 3DyRM parameters.
Thread migration is then used to modify the state of each thread to simultaneously optimise the parameters.
However, GIPS, intsB and latency have values with different orders of magnitude.
For this situation, weighting methods are recommended to aggregate the parameters~\cite{Cheng2002}.
Therefore, we propose to characterise each thread using an aggregate objective function, $P$, that combines the three parameters.

Let $P_{ijk}$ be the performance for the $i$-th thread of the $j$-th process when executed on the $k$-th of $N$ nodes.
Then, for each iteration of the aggregate function,

\begin{equation}
\label{eq:p_1}
P_{ijk}=\frac{\mathrm{GIPS}_{ijk}^\beta \cdot \textrm{intsB}_{ijk}^\gamma}{\textrm{latency}_{ijk}^\alpha}
\end{equation}

\noindent where $\textrm{GIPS}_{ijk}^\beta$ is the GIPS of the thread powered by $\beta$, and
$\textrm{instB}_{ijk}^\gamma$ and ${\textrm{latency}_{ijk}^\alpha}$ are the instB and latency values powered by $\gamma$ and $\alpha$, respectively. It is clear that larger values of $P_{ijk}$ imply better performance.

Initially, no values of $P_{ijk}$ are available for any thread on any node.
On each time interval, $P_{ijk}$ is computed for every thread on the system according to the performance read by the hardware counters.
In every interval some values of $P_{ijk}$ are updated, for those nodes $k$ where each thread was executed, while others store the performance information of each thread when it was executed in a different node (if available).
If there is a previous value of $P_{ijk}$, the new value replaces the previously saved one.
Thus, the algorithm adapts to possible behaviour changes for the threads.
For example, in a Xeon server with four nodes, $N=4$, four values of $P$ (one for each thread) are saved each iteration.
As threads migrate and are executed on different nodes, $P_{ijk}$ are progressively updated.

Every $T$ milliseconds, once the new values of $P_{ijk}$ are computed,  the thread with the worst current performance, in terms of $P_{ijk}$, is selected to be migrated.
To compare threads from different processes, each individual $P_{ijk}$is divided by the mean $P_{ijk}$ of all threads of the same process (same $j$), identified by its PID, 

\begin{equation}
\label{eq:p_3}
\widehat{P}_{ijk}= \frac {P_{ijk}}{\sum_{m=1}^{n_j} P_{mjh} / n_j}
\end{equation}

\noindent where $n_j$ is the number of threads of process $j$ and $h$ is, for each thread $m$ of process $j$, the last node where it was executed.
Thus, for each process, those threads with $\widehat{P}_{ijk}<1$ are currently performing worse than the mean of the threads in the same process, and the worst performing thread in the system is considered to be the one with the lowest $\widehat{P}_{ijk}$, i.e., the thread performing worse when compared to the other threads of its process.
This is the migration thread, denoted by $\Theta_m$.

The migration can be to any core in a node other than $\Theta_m$ current node.
A weighted random process is employed to choose the destination, based on the stored performance values.
The aim is to consider all possible migrations, and so all $P_{ijk}$ values are updated and behavioural changes are incorporated.

In order to consider all possible migrations, all $P_{ijk}$ values are important.
Therefore, one of the aims is to fill as many entries of $P_{ijk}$ as possible.
To ensure the migration will improve performance, every possible destination is granted a number of tickets according to the likelihood of that migration improving performance, and the destination with the larger likelihood overall is chosen.
Migration may take place to an empty core, where no other thread is currently being executed, or to a core occupied with other threads.
If there are already threads in the core, one would have to be exchanged with $\Theta_m$.
The swap thread is denoted as $\Theta_g$, and all threads are candidates to be $\Theta_g$.
Note that, although not all threads may be selected to be $\Theta_m$, e.g.\ a process with a single thread would always have $\widehat{P}_{ijk}=1$ and so never be selected, they may still be considered to be $\Theta_g$ to ensure we obtain the best possible performance for the whole system.

The rules applied to distribute tickets ($B$) for the random selection procedure are:
\begin{itemize}
\item Destinations in nodes where $\Theta_m$ has previously performed worse than the current node get $B_1$ tickets.
\item Destinations in nodes where there is no previous data recorded for $\Theta_m$ get $B_2$ tickets.
\item Destinations in nodes where $\Theta_m$ has previously performed better than the current node get $B_3$ tickets. 
\end{itemize}

The best migration should be that which results in good performance from both threads, $\Theta_m$ and $\Theta_g$.
Therefore, additional tickets are awarded to each destination according to the values of $P_{gn}$, where $g$ is $\Theta_g$, and $n$ is the node that currently hosts $\Theta_m$: 
\begin{itemize}
\item Destinations where $\Theta_g$ has previously performed worse in $n$ in the past get $B_4$ tickets.
\item Destinations with no previous information for $\Theta_g$ get $B_5$ tickets.
\item Destinations where $\Theta_g$ has previously performed better get $B_6$ tickets.
\item Destinations for cores with no threads assigned get $B_7$ tickets.
\end{itemize}  

Although $P_{ijk}$ are only saved for nodes, by including the performance of the possible $\Theta_g$, different cores in the same node, and even different threads in the same core, may get a different number of tickets.

Clearly, suitable choice of $B_k$ is critical, and this is discussed further below.

When all tickets have been assigned, a final destination core is randomly selected based on the awarded tickets.
The interchanging thread, $\Theta_g$, is  chosen from those currently being executed on that core, if the core is not free.
Once the threads to be migrated are selected, the migrations are actually performed.

This algorithm is referred to as the interchange migration algorithm with performance record (IMAR).
To simplify notation, an IMAR and parameters is denoted as IMAR[$T;\alpha,\beta,\gamma$].

However, migrations may affect not only the involved threads, $\Theta_m$ and $\Theta_g$, but all threads in the system due to synchronisation or other collateral relations among threads.
These relations are not accurately modelled using each thread performance separately.
Therefore, we propose the interchange algorithm with performance record and rollback (IMAR\textsuperscript{2}), where the total performance for each iteration is calculated as the sum of all $P_{ijk}$ for all threads.
Thus, the current total performance, $Pt_{current}$,  a single value, is available to evaluate a thread configuration, independent of the processes being executed.
The total performance of the previous iteration is stored as $Pt_{last}$.

Incorporating these concepts, decisions are made regarding the next iterations of the algorithm.
The algorithm may dynamically adjust the rate of migrations by altering $T$ between a given minimum, $T_{min}$, and maximum, $T_{max}$, doubling or halving the previous value.
To do that, a ratio,  $0< \omega\le 1$ is defined for $Pt_{current} / Pt_{last}$, to limit an acceptable decrement in performance.
So, if a thread placement has low total performance, migrations should be performed to obtain better thread placement, because they are likely to increase performance ($Pt_{current} \ge \omega Pt_{last}$). This way, $T$ is decreased to perform migrations more often and reach optimal placement quicker.
On the other hand, if thread placement has high total performance, migrations have a greater chance of being detrimental.
In this case, if $Pt_{current} < \omega Pt_{last}$, there is no requirement for many migrations, so $T$ is increased.
Additionally, a rollback mechanism is implemented, to undo migrations if they result in a significant loss of performance, returning migrated threads  to their former locations. If a rollback is performed, no other migrations are made during that interval.

Summarising, the rules guiding our algorithm are:
\begin{itemize}
\item{If $Pt_{current} >= \omega Pt_{last}$}, i.e., the total performance improves: Migrations are considered productive,  $T$ is halved ($T=>T/2$), and a new migration is performed according to IMAR.
\item{If $Pt_{current} < \omega Pt_{last}$}, i.e., the total performance decreases more than a given threshold: Migrations are considered counter-productive, $T$ is doubled ($T=>2 \times T$), and the last migration is rolled back.
\end{itemize}

The algorithm continues to migrate threads to allow for changes in system behaviour, and to obtain performance information, rolling these back if necessary (rollback).
To simplify notation,  IMAR\textsuperscript{2} and parameters is denoted as IMAR\textsuperscript{2}[$T_{min},T_{max};\alpha,\beta,\gamma;\omega$].

A simple example is presented to clarify our proposal.
Consider a system with 6 cores in three different nodes, incorporating  three processes, each with two threads.
Initially, Process 1 ($PID=j=100$) has 2 threads ($i=1,2$, $TID=100,101$) executed in node 0 (cores 0 and 1), process 2 ($PID=j=200$) has 2 threads ($i=1,2$, $TID=200,201$) executed in node 1 (cores 2 and 3), and Process 3 ($PID=j=300$) has 2 threads ($i=1,2$, $TID=300,301$) executed in node 2 (cores 4 and 5), as shown in Table~\ref{table_example_IMAR_1}, where threads are shown with the core they currently reside, and their recorded performance in each node.
Nodes where threads have not been executed previously have no performance information recorded.

Table~\ref{table_example_IMAR_2} shows a later state, where some migrations have been executed and more performance information is available. The performance of each thread in its current node is shown in bold.
Suppose a migration has to be decided.
Table~\ref{table_example_IMAR_3} shows each thread’s performance and normalised performance ($\widehat{P}_{ijk}$, $P_{ijk}$ divided by the current mean performance of all the threads in the same process (same $j$), equation~\ref{eq:p_3}).
Thread 300 has the worst relative performance, so $\Theta_m$=300.

The case studies, Section~\ref{sec:CasesStudy}, show optimal values for $B_k$ to be 
\begin{itemize}
\item $B_1=B_4=1$: previous low performances are penalised,
\item $B_2=B_5=2$: allow more performance information to be obtained,
\item $B_3=B_6=4$: previous good performances are rewarded, and
\item $B_7=3$: allow migrations to free cores and improve load balance.
\end{itemize}
With these values, a thread interchange that would increase the performance of both threads involved would get eight tickets, the maximum, whereas one that would worsen the performance of both threads would get only two tickets, the minimum, and, therefore, have 1/4 the chance of being selected.
Migrations and interchanges where there are no data still have a chance of being selected, providing (eventually) values for all possible $P_{ijk}$.

 \begin{table}[tpb]
\renewcommand{\arraystretch}{1.3}
\caption{Example IMAR: Initial state.}
\label{table_example_IMAR_1}
\centering
\begin{tabular}{|c|c||c|c|c|}
\hline
TID (i,j) & core & $P_{ij0}$ & $P_{ij1}$ & $P_{ij2}$\\
\hline
\hline
100 (0,100)& 0 & \textbf{2.4} & -- & --\\
101 (1,100)& 1 & \textbf{2.6} & -- & --\\
\hline
200 (0,200)& 2 & -- & \textbf{1.4} & -- \\
201 (1,200)& 3 & -- & \textbf{1.6} & -- \\
\hline
300 (0,300)& 4 & -- & -- & \textbf{6.3} \\
301 (1,300)& 5 & -- & -- & \textbf{5.2} \\
\hline
\end{tabular}
\end{table}

 \begin{table}[tbp]
\renewcommand{\arraystretch}{1.3}
\caption{Example IMAR: State after a number of iterations.}
\label{table_example_IMAR_2}
\centering
\begin{tabular}{|c|c||c|c|c|}
\hline
TID (i,j) & core & $P_{ij0}$ & $P_{ij1}$ & $P_{ij2}$\\
\hline
\hline
100 (0,100)&2 & 2.5 & \textbf{1.9} & 2.9\\
101 (1,100)&4 & 2.7 & 1.8 & \textbf{3.1}\\
\hline
200 (0,200)&0 & \textbf{0.9} & 1.4 & --\\
201 (1,200)&5 & -- & 1.6 & \textbf{2.1}\\
\hline
300 (0,300)&1 & \textbf{3.3} & -- & 6.3 \\
301 (1,300)&3 & -- & \textbf{8.1} & 5.7 \\
\hline
\end{tabular}
\end{table}

 \begin{table}[tbp]
\renewcommand{\arraystretch}{1.3}
\caption{Thread performance for the example of Table II.}
\label{table_example_IMAR_3}
\centering
\begin{tabular}{|c||c|c|c|c|c|c|}
\hline
 -- & 100 & 101 & 200 & 201 & 300 & 301\\
\hline
\hline
$P$ & 1.9 & 3.1 & \textbf{0.9} & 2.1 & 3.3 & 8.1\\
 $\widehat{P}$ & 0.76 & 1.24 & 0.6 & 1.4 & \textbf{0.58} & 1.42\\
\hline
\end{tabular}
\end{table}

Table~\ref{table_example_IMAR_4} shows the distribution of tickets for this example, where destinations can be considered the same as cores or threads, because there is only one thread per core and no idle cores.
Tickets are awarded according to the past performance of thread $\Theta_g$=300.
\begin{itemize}
\item Thread 300 cannot move to core 1 (its current location) or 0 (in the same node), so both get 0 tickets.
\item Cores 2 and 3 get $B_2$ tickets, since there is no past information of thread 300 on node 1. 
\item Cores 4 and 5 get $B_3$ tickets, because performance of thread 300 was better on node 2 than on the current node. 
\end{itemize}
Tickets are then awarded considering the past performance of the threads that are currently executing on each particular core, when executed previously on node 0, the node currently hosting thread 300.
\begin{itemize}
\item Core 2 gets $B_6$ tickets because thread 100 performed better on node 0.
\item Core 3 gets $B_5$ tickets because thread 301 has no previous performance information on node 0.
\item Core 4 gets $B_4$ tickets because thread 101 performed worse on node 0.
\item Core 5 gets $B_4$ tickets because thread 201 has no previous performance information on node 0.
\end{itemize}
Thus, 21 tickets were awarded in total, and
\begin{itemize}
\item Thread 300 has 6/21 chances of migrating to core 2 and being interchanged with thread 100. This would be favourable to thread 100 and unknown to thread 300.
\item Thread 300 has 6/21 chances of moving to core 5 and being interchanged with thread 201. This would be unknown to thread 201 and favourable to thread 300.
\item Thread 300 has 5/21 chances of migrating to core 4 and being interchanged with thread 101. This would be detrimental to thread 101 and favourable to thread 300.
\item Thread 300 has 4/21 chances of going to core 3 and being interchanged with thread 301. This would be detrimental to thread 301 and unknown to thread 300.
\end{itemize}

Once all tickets are awarded, $\Theta_g$ is chosen in a lottery.
The interchange can be performed when $\Theta_m$ and $\Theta_g$ are chosen, migrating both threads to each other cores.
Note that this is a small example, in a real situation with more threads and nodes, the probability differences among the possible migrations would be larger.

 \begin{table}[tbp]
\renewcommand{\arraystretch}{1.3}
\caption{Ticket distribution for the example of Table II.}
\label{table_example_IMAR_4}
\centering
\begin{tabular}{|c|c||c|c|c||c|}
\hline
TID (i,j) & core & $P_{ij0}$ & $P_{ij1}$ & $P_{ij2}$ & tickets\\
\hline
\hline
100 (0,100)&2 & 2.5 & \textbf{1.9} & 2.9 & $B_2$ + $B_6$ = 2+4\\
101 (1,100)&4 & 2.7 & 1.8 & \textbf{3.1} & $B_3$ +$B_4$ = 4+1\\
\hline
200 (0,200)&0 & \textbf{0.9} & 1.4 & -- & 0\\
201 (1,200)&5 & -- & 1.6 & \textbf{2.1} & $B_3$ + $B_5$= 4+2\\
\hline
\textbf{300} (0,300)&1 & \textbf{3.3} & -- & 6.3 & 0\\
301 (1,300)&3 & -- & \textbf{8.1} & 5.7 & $B_2$ + $B_5$=2+2\\
\hline
\end{tabular}
\end{table}

\section{Experimental results}
\label{sec:CasesStudy}
NPB-OMP benchmarks were used to study the effect of the memory allocation.
These benchmarks are well suited for multicore processors, although they do not greatly stress the memory of large servers.
To simulate the effects of NUMA memory allocation, different memory stress situations were simulated using the numactl tool~\cite{kleen2005numa}, which allows the memory cell to store data can to be defined and threads to be pinned to specific cores or processors.
We designed an experiment where four instances of the NPB-OMP benchmarks are executed concurrently in a multiprocessor system, and the placement of each could be controlled.
Each benchmark instance was executed in one multi-threaded process.
The system employed was an Ubuntu~14, with Linux kernel 3.10, NUMA server with four nodes, each of which had one octo-core Xeon E5-4620 (32 physical cores in total), Sandy Bridge architecture, 16 MB L3 cache, 2.2 GHz-2.6 GHz, and 512 GB of RAM.
Node 0 contained cores 0 to 7, node 1 contained cores 8 to 15, node 2 contained cores 16 to 23, and node 3 contained cores 24 to 31.
Each benchmark was executed with just enough threads to fill one node.
Thus, each process could have its execution threads pinned to any node and its data assigned to a selected memory cell.
Different memory placement scenarios could be established by executing as many process as nodes.
We tested the options:
\begin{itemize}
 \item \textsc{free} test: The benchmarks started execution at the same time, and the OS controlled memory and thread placement.
 \item \textsc{direct} test: Each benchmark had its threads fixed to one node and preferred memory set to the same cell.
 \item \textsc{crossed} test: Each benchmark had its threads fixed to one processor and preferred memory set to a different cell. When more than two cells were considered, there were several possible combinations. The configuration used in the case study with four cells was:
\begin{itemize}
\item threads in node 0 had their data in cell 1, 
\item threads in node 1 had their data in cell 0, 
\item threads in node 2 had their data in cell 3, and 
\item threads in node 3 had their data in cell 2.
\end{itemize}
 \item \textsc{interleave} test: Each benchmark had its threads fixed to one node and memory set to interleave, with each consecutive memory page set to a different memory cell in a round robin fashion.
\end{itemize}

Four class C NPB-OMP codes were selected to be shown here: \texttt{lu.C}, \texttt{sp.C}, \texttt{bt.C} and \texttt{ua.C}.
This selection was made according to two main criteria:
\begin{itemize}
\item Codes with different memory access patterns and different computing requirements.
The DyRM model  
was used to select two benchmarks with low flopsB (\texttt{lu.C} and \texttt{sp.C}) and two with high flopsB (\texttt{bt.C} and \texttt{ua.C}).
\item Execution time.
Since the execution times of these codes are similar, they remain in concurrent execution most of the time.
This helps studying the effect of thread migrations.
\item They are representative of the benefits obtained by our proposal and other experiments do not show different behaviours.
\end{itemize}
All benchmarks were compiled with gcc and O2 optimisation.

 \subsection{Baseline results}
 \label{ssec:baseline}
The effects of the memory placements in the execution of the NPB-OMP benchmarks are evaluated, with threads pinned to the the same core for the duration of their execution, so no migrations take place.
These results are used as a baseline to evaluate IMAR and IMAR\textsuperscript{2}.
Each test was executed on the four nodes, combined as four processes of the same code that produced four combinations (\textbf{4 lu.C}, \textbf{4 sp.C}, \textbf{4 bt.C}, and \textbf{4 ua.C}), and four processes of different codes, that produced one combination (\textbf{lu.C/sp.C/bt.C/ua.C}).
Every test was executed five times and the mean execution times are shown in Table~\ref{table_baseline_4cell}. The times for all benchmarks of \textbf{lu.C/sp.C/bt.C/ua.C} are shown, whereas, for considerations of space,   only the times of the fastest and slowest instances are shown for the four equal benchmarks. 

Since the NPB-OMP benchmarks perform reasonably well on multicores and they do not stress the memory, the \textsc{free} test, where the OS placed threads and memory, performs reasonably well, although inferior to the \textsc{direct} case.
\texttt{sp.C} is inferior to the \textsc{direct} test, but only when executed with other codes in the \textbf{lu.C/sp.C/bt.C/ua.C} combination.
For that case, when benchmarks \texttt{ua.C} and \texttt{bt.C} finish execution in the \textsc{free} test, the OS is free to place \texttt{sp.C} threads in other processors to balance the load, which leads to a faster execution.
\texttt{sp.C} is memory intensive, and the OS accelerates its execution by making the opposite application slower.
For the other cases, placing memory and execution threads in the same node appears to be the best option, interleaving memory does not produce good results.
As expected, by far most inferior case is the \textsc{crossed} test, where memory and threads are on different nodes.

Thus, the \textsc{direct} case is the best option, although it has some load balancing issues that can decrease the global performance, and the \textsc{crossed} test is the worst case, as expected.

 \begin{table}[tbp]
\renewcommand{\arraystretch}{1.3}
\caption{Baseline times for four NAS benchmarks.}
\label{table_baseline_4cell}
\centering
\begin{tabular}{|c|c||c|c|c|c|}
\cline{1-6}
 \multicolumn{2}{|c||}{Test} & \multicolumn{4}{|c|}{Time (s)}\\
\hline
 concurrent benchmarks & benchmark & \textsc{free} & \textsc{direct} & \textsc{interleave} & \textsc{crossed} \\
\hline
\hline
\hline
\multirow{4}{*}{lu.C/sp.C/bt.C/ua.C} & lu.C & 220.24 & 210.00 & 428.41& 1221.05\\
& sp.C & 235.53 & 267.89	& 557.39	& 1698.36\\
& bt.C & 201.69 & 180.77	& 260.46	& 500.037\\
& ua.C & 197.03 & 190.26	& 316.26	& 759.17\\
\hline
\hline
\multirow{2}{*}{4 lu.C} & fastest lu.C & 213.09 & 209.99 & 444.09 & 1265.46\\
& slowest lu.C & 215.84 & 212.20 & 452.15 & 1278.86\\
\hline
\multirow{2}{*}{4 sp.C} & fastest sp.C & 267.80 & 265.29 & 511.15 & 1848.41\\
& slowest sp.C & 287.49 & 267.71 & 763.88 & 1864\\
\hline
\multirow{2}{*}{4 bt.C} & fastest bt.C & 181.27 & 180.74 & 242.52 & 452.47\\
& slowest bt.C & 185.37 & 182.29 & 246.90 & 453.13\\
\hline
\multirow{2}{*}{4 ua.C} & fastest ua.C &194.51 & 189.36 & 303.76 & 677.31\\
& slowest ua.C & 203.54 & 190.46 & 313.59 & 684.70\\
\hline
\end{tabular}
\end{table}

\subsection{Traces}
\begin{figure*}[tbp]
 \begin{center}
     \includegraphics[width=.95\textwidth]{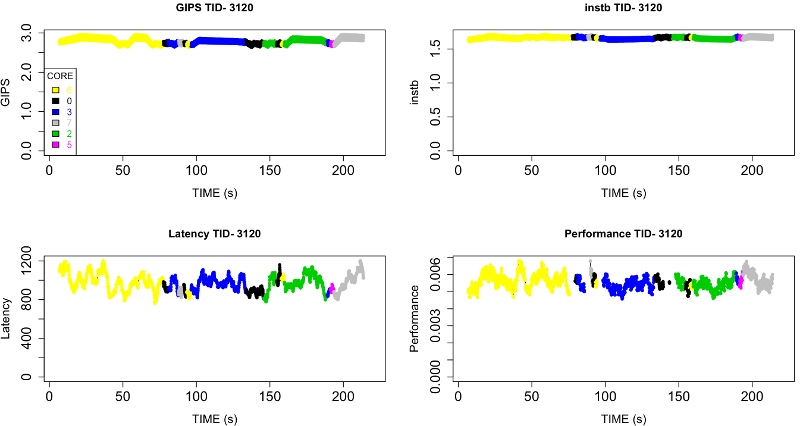}
 \end{center}
 \caption{Evolution of performance for one thread of the \textbf{4 lu.C} configuration for the \textsc{direct} case. The thread runs in node 0.}
 \label{fig:4lu_perf_direct}
 \end{figure*}

\begin{figure*}[tbp]
 \begin{center}
     \includegraphics[width=.95\textwidth]{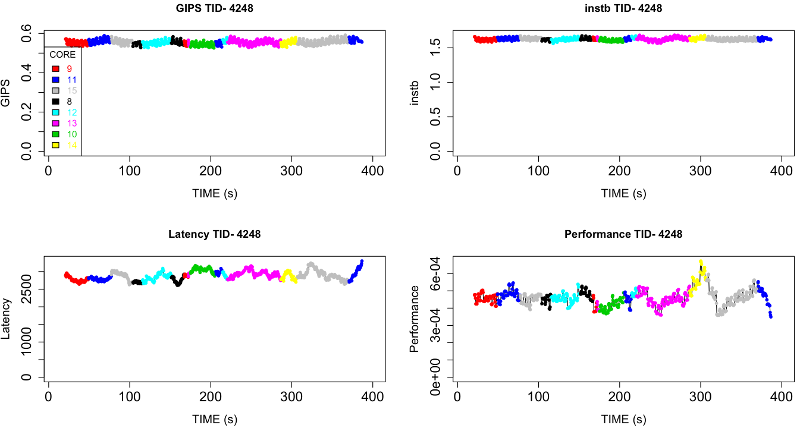}
 \end{center}
 \caption{Evolution of performance for one thread of the \textbf{4 lu.C} configuration for the \textsc{crossed} case. The thread runs in node 1.}
 \label{fig:4lu_perf_crossed}
 \end{figure*}

The migration tool can be configured to dump the PEBS trace to a file, which can be read by a performance visualisation tool, such as in~\cite{lorenzo2014}.
Thus, the evolution of the performance of each thread, in terms of $P_{ijk}$ and its components, through the execution of the benchmarks can be plotted.
Figures~\ref{fig:4lu_perf_direct} and~\ref{fig:4lu_perf_crossed} show the performance of a thread of the \textbf{4 lu.C} benchmark in the \textsc{direct} and \textsc{crossed} configurations, respectively.
These figures show the evolution in time of $P_{ijk}$ for a given thread and each of its performance components, GIPS, instB, and latency, from eq.~\ref{eq:p_1}.
Different line colours represent different cores, and a change in colour represents a migration of the thread.
To better visualise the changes, we used a frame average of 50 measurements, corresponding to measurement every 1.5 seconds.
This frame average implies that performance changes between migrations are not instantly visible, but usually take the form of peaks and valleys.
In Figures~\ref{fig:4lu_perf_direct} and~\ref{fig:4lu_perf_crossed}, migrations were performed by the OS among cores in the same node, so performance does not vary greatly during execution.
As expected, performance is lower on the \textsc{crossed} test, with more migrations involved.
 
         \begin{figure*}[tbp]
 \begin{center}
     \includegraphics[width=.95\textwidth]{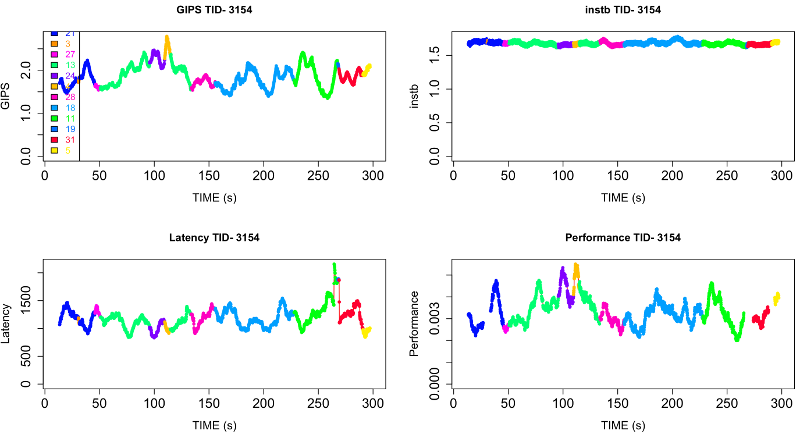}
 \end{center}
 \caption{Evolution of performance for one thread of the \textbf{4 lu.C} configuration for the \textsc{crossed} test with IMAR migrations.}
 \label{fig:4lu_perf_crossed_migs_3}
 \end{figure*} 
 
Figure~\ref{fig:4lu_perf_crossed_migs_3} shows the performance of a thread during the execution of the \textbf{4 lu.C} combination in the \textsc{crossed} test employing IMAR migrations.
Performance increases and approaches the \textsc{direct} case due to the IMAR migrations.
Using IMAR, migrations take place between nodes, so they influence the performance more than in Figure~\ref{fig:4lu_perf_crossed}.
Peaks in the graph of the same colour, are likely due to migrations of other threads that influence the single-colour thread, whereas peaks with a colour change are due to migrations of the thread itself.
Note that migrations usually occur after a performance dip, because the thread was chosen to be among the worst performing by the IMAR algorithm. 
For example, the migration after 250 seconds is apparently due to an increase in memory latency.

        \begin{figure*}[tbp]
 \begin{center}
     \includegraphics[width=.95\textwidth]{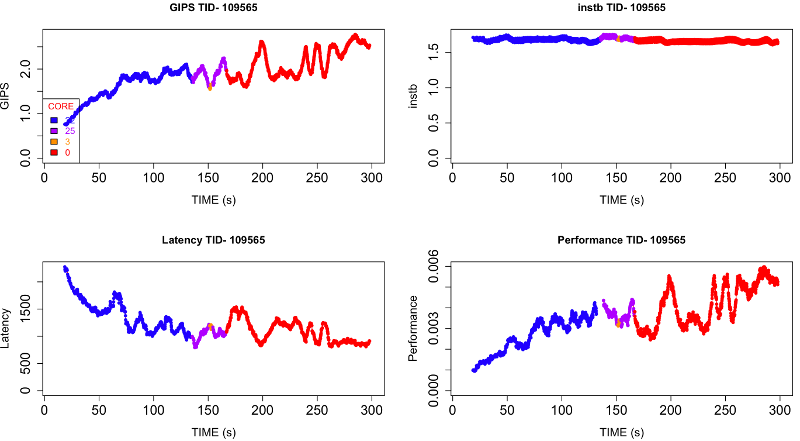}
 \end{center}
 \caption{Evolution of performance for one thread of the \textbf{4 lu.C} configuration for the \textsc{crossed} test with IMAR\textsuperscript{2} migrations.}
 \label{fig:4lu_perf_crossed_migs_097_1}
 \end{figure*} 
 
         \begin{figure*}[tbp]
 \begin{center}
     \includegraphics[width=.95\textwidth]{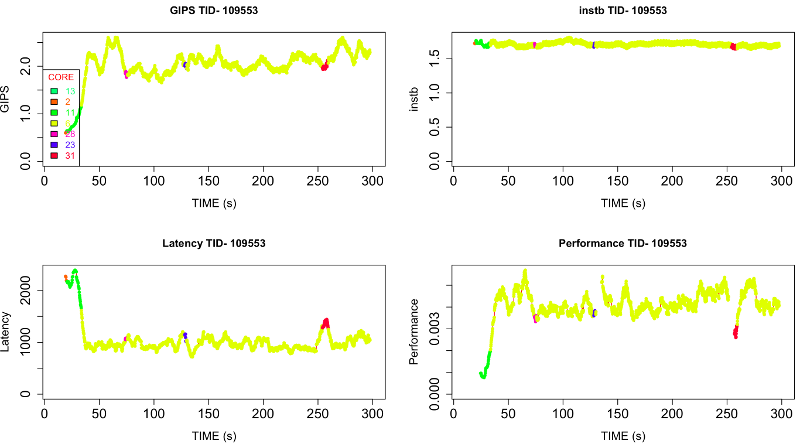}
 \end{center}
 \caption{Evolution of performance for one thread of the \textbf{4 lu.C} configuration for the \textsc{crossed} test with IMAR\textsuperscript{2} migrations.}
 \label{fig:4lu_perf_crossed_migs_097_3}
 \end{figure*} 

The performance of two threads during the execution of the \textbf{4 lu.C} combination in the \textsc{crossed} test and IMAR\textsuperscript{2} migrations ($\omega=0.97$) are shown in Figures~\ref{fig:4lu_perf_crossed_migs_097_1} and~\ref{fig:4lu_perf_crossed_migs_097_3}.
The tendency towards increasing performance is clear, because the rollbacks reduce the number of migrations.
There are less pronounced variations in performance than in the IMAR case, due varying $T$ and rollback strategies.
A dip in performance of thread 109565 close to 150 seconds (Fig.~\ref{fig:4lu_perf_crossed_migs_097_1}) triggers a migration from core 3 back to core 25, a rollback.
A migration from core 13 (in node 1) places the thread in core 6 (in node 0) (Fig.~\ref{fig:4lu_perf_crossed_migs_097_3}), and subsequently there are   rollbacks around 70, 130, and 260 seconds, which indicate that thread 109553 was placed in an efficient node, and it is inefficient to move it.
The IMAR\textsuperscript{2} algorithm explores all possible placements for all threads, and so counter-productive migrations can be performed, but including rollback allows their effects to be minimised. 
 The algorithm tries other node placements for the thread, computing the whole performance record (moving to core 28, node 4, to core 23, node 3, etc.) and checking for behaviour changes, but always returns the thread to core 6 in node 0.
 
          \begin{figure*}[tbp]
 \begin{center}
     \includegraphics[width=\textwidth]{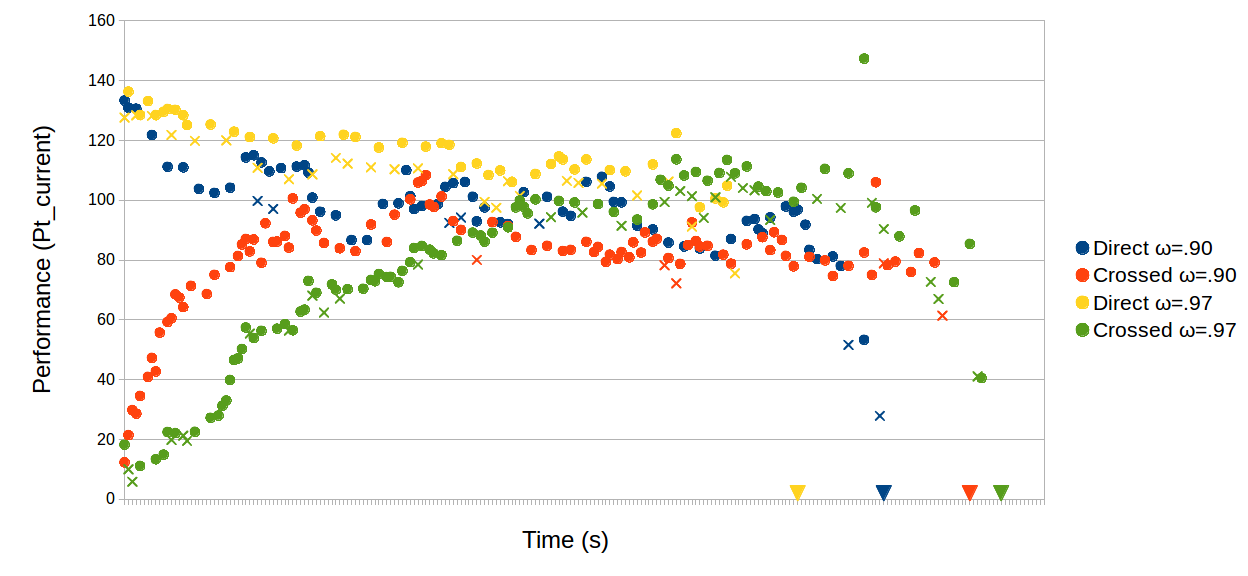}
 \end{center}
 \caption{Evolution of performance for the \textbf{4 lu.C} configuration for the \textsc{crossed} and \textsc{direct} cases with IMAR\textsuperscript{2}[1,4;1,1,1;0.90] and IMAR\textsuperscript{2}[1,4;1,1,1;0.97].}
 \label{fig:perf_graph}
 \end{figure*} 

An example of migration timing in the \textbf{4 lu.C} combination for the \textsc{crossed} and \textsc{direct} tests with IMAR\textsuperscript{2} migrations is shown in Figure~\ref{fig:perf_graph}.
Thresholds $\omega=0.90$ and $\omega=0.97$ were considered, and the performance record for the whole system is shown, where
a circle represents a migration, a cross represents a rollback, and
triangles mark the execution time of each test.
This graph is from a single execution of each case for each value of $\omega$.

In \textsc{direct} cases, performance remains higher with $\omega=0.97$ through the executions, due to rollbacks, since migrations are counter-productive in this case.
When performance dips, a rollback is executed (a yellow cross in the figure) and it recovers.

In the \textsc{crossed} configurations, where migrations are initially productive, when all threads are in inefficient placements, performance increases faster with $\omega=0.90$ than with $\omega=0.97$.
With $\omega=0.90$, no rollbacks are performed during the first minute, while rollbacks with $\omega=0.97$ are counter-productive, since they make the process slower when approaching the best placements.
Nevertheless, once performance is high enough, and more threads are correctly placed, the $\omega=0.97$ case helps keep performance high with rollbacks, whereas when $\omega=0.90$, migrations continue even once a good configuration is obtained.

\subsection{Results for IMAR}
\begin{figure*}[tbp]
 \begin{center}
     \includegraphics[width=\textwidth]{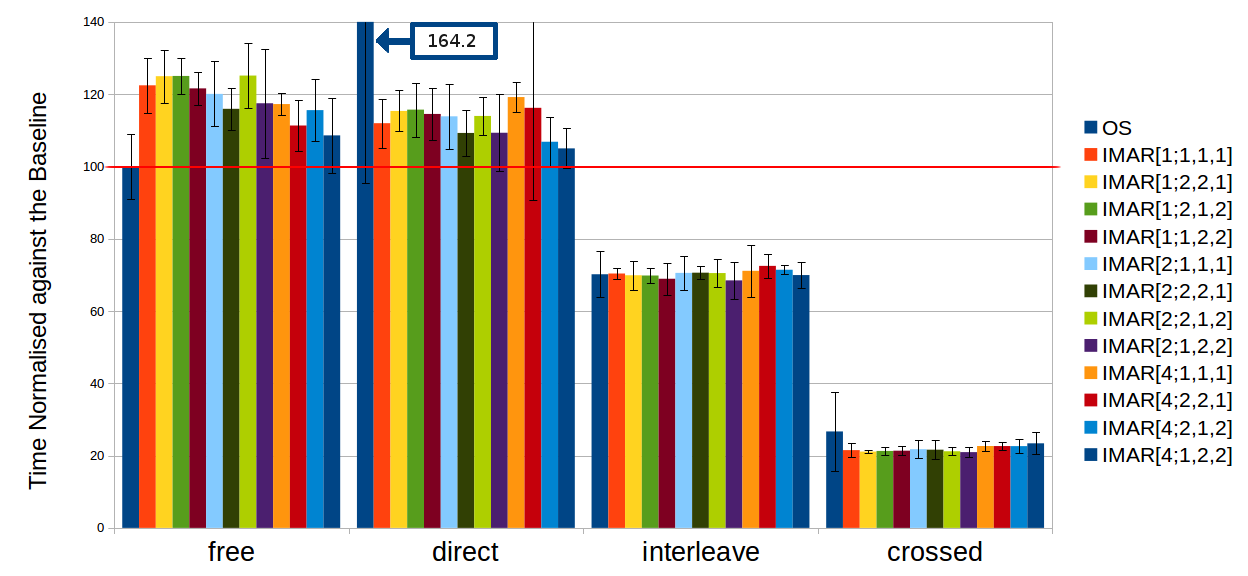}
 \end{center}
 \caption{Normalised execution times for lu.C in the \textbf{lu.C/sp.C/bt.C/ua.C} test with IMAR.}
 \label{fig:4multi_luC}
 \end{figure*}

\begin{figure*}[tbp]
 \begin{center}
     \includegraphics[width=\textwidth]{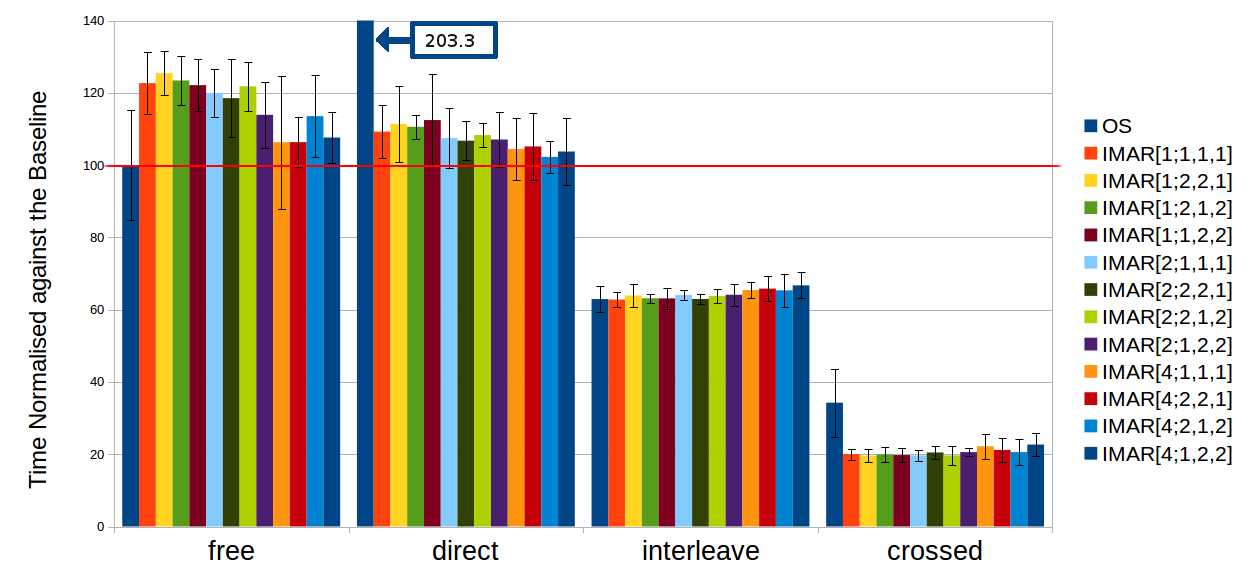}
 \end{center}
 \caption{Normalised execution times for sp.C in the \textbf{lu.C/sp.C/bt.C/ua.C} test with IMAR.}
 \label{fig:4multi_spC}
 \end{figure*}

\begin{figure*}[tbp]
 \begin{center}
     \includegraphics[width=\textwidth]{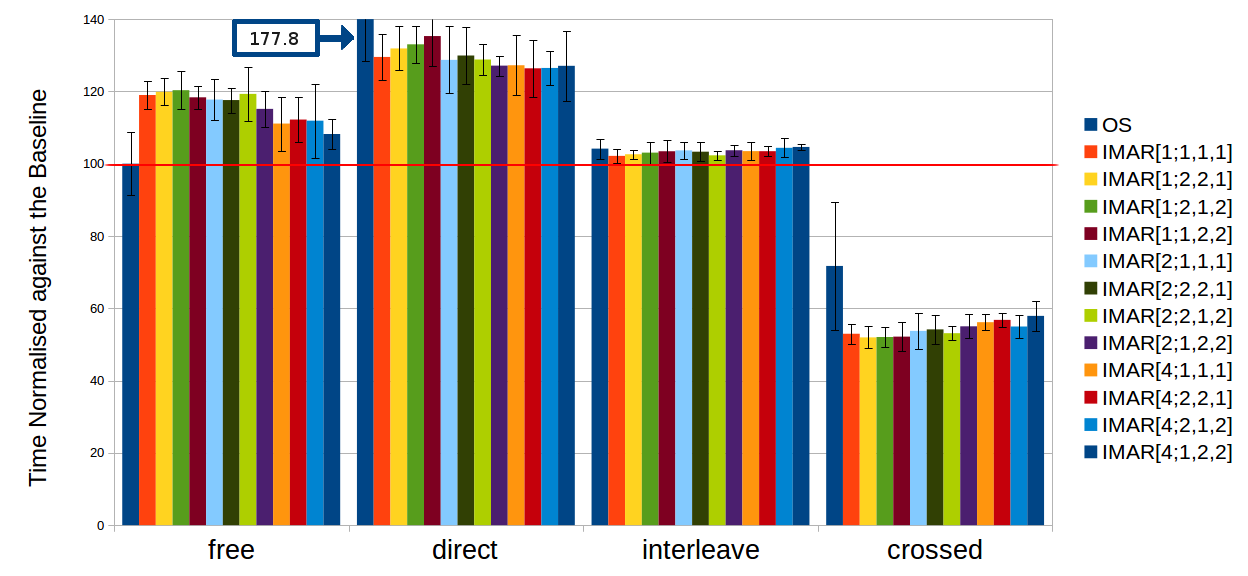}
 \end{center}
 \caption{Normalised execution times for bt.C in the \textbf{lu.C/sp.C/bt.C/ua.C} test with IMAR.}
 \label{fig:4multi_btC}
 \end{figure*}

\begin{figure*}[tbp]
 \begin{center}
     \includegraphics[width=\textwidth]{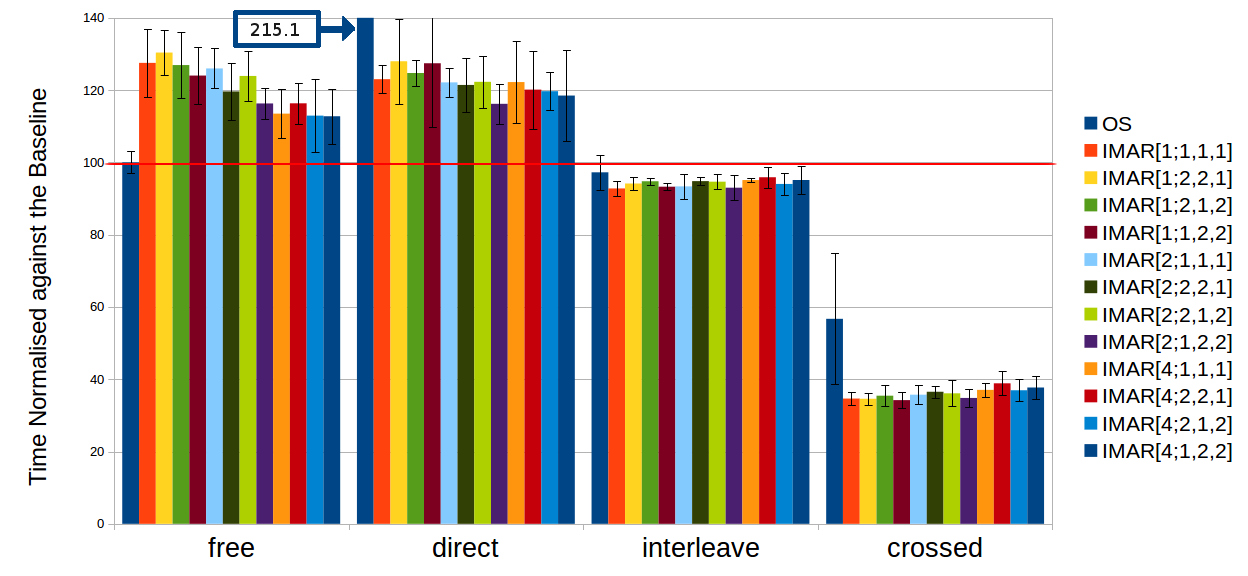}
 \end{center}
 \caption{Normalised execution times for ua.C in the \textbf{lu.C/sp.C/bt.C/ua.C} test with IMAR.}
 \label{fig:4multi_uaC}
 \end{figure*}

We discuss variations in execution time of our tests compared to the baseline results of Table~\ref{table_baseline_4cell}.
Figures~\ref{fig:4multi_luC}--\ref{fig:4multi_uaC} show the results of one benchmark for all the tests with one combination.
Executions using IMAR with different values of $T$, $\alpha$, $\beta$, and $\gamma$ are also shown.

All figures in this next sections show the experimental execution times performing migrations, by IMAR, as a proportion of the baselines times of each test (\textsc{free}, \textsc{direct}, \textsc{interleave} and \textsc{crossed}), expressed as a percentage.
A percentage greater that 100 means a worse execution time, while a result under 100 shows a better execution time.
A special case is shown for the OS, where the \textsc{direct}, \textsc{interleave} and \textsc{crossed} tests are modified to fix only the memory placement, letting the OS select thread placement.
These tests show whether the OS was able to place the threads near the memory where their data are stored.
Note that, in many of the figures, the bar for the OS result in the \textsc{direct} case is far higher than the rest, so it was cropped and the actual result is shown in a box by the bar.

In this case, the effect of $T$, which determines the number of migrations, is critical.
On most of these tests, the benchmarks use the same code, which makes comparing their performance fairer and easier.
For the \textbf{lu.C/sp.C/bt.C/ua.C} combination, there is an apparent bias towards applications with low instB (Figs.~\ref{fig:4multi_luC},~\ref{fig:4multi_spC},~\ref{fig:4multi_btC} and~\ref{fig:4multi_uaC}), with superior results for \textbf{lu.C} and \textbf{sp.C} than for \textbf{bt.C} and \textbf{ua.C}.
Note that \textbf{bt.C} and \textbf{ua.C}  execute faster, and so must always share the system among four benchmarks, whereas \textbf{lu.C} and \textbf{sp.C} have more free cores at the end of their execution.
 This situation produces superior performance, in part due to frequency scaling capabilities on Xeon systems, when core frequency increases if not all cores are active.

Changing the scaling factors $\alpha$, $\beta$, and $\gamma$ has a slight impact on the effect of the migrations.
For example, consider the \textbf{lu.C/sp.C/bt.C/ua.C} combination.
For  \textbf{lu.C}, Fig.~\ref{fig:4multi_luC}, configurations which give greater importance to memory latency, IMAR[$T;2,2,1$] and IMAR[$T;2,1,2$], for $T=1$, $T=2$ and $T=4$, are superior in the \textsc{direct} and \textsc{crossed} tests, where data locality is more important, and inferior in the \textsc{interleave} test, where memory latency is more balanced in all nodes.
Figure~\ref{fig:4multi_spC}, corresponding to \textbf{sp.C}, shows similar outcomes to \textbf{lu.C}, since they are both memory intensive benchmarks, but with more clear influence of the migrations, since memory latency is more important.
Figures~\ref{fig:4multi_btC} and~\ref{fig:4multi_uaC} show less difference among configurations because latency is not so important in these cases.

\subsection{Results for IMAR\textsuperscript{2}}
\begin{figure*}[tbp]
 \begin{center}
     \includegraphics[width=\textwidth]{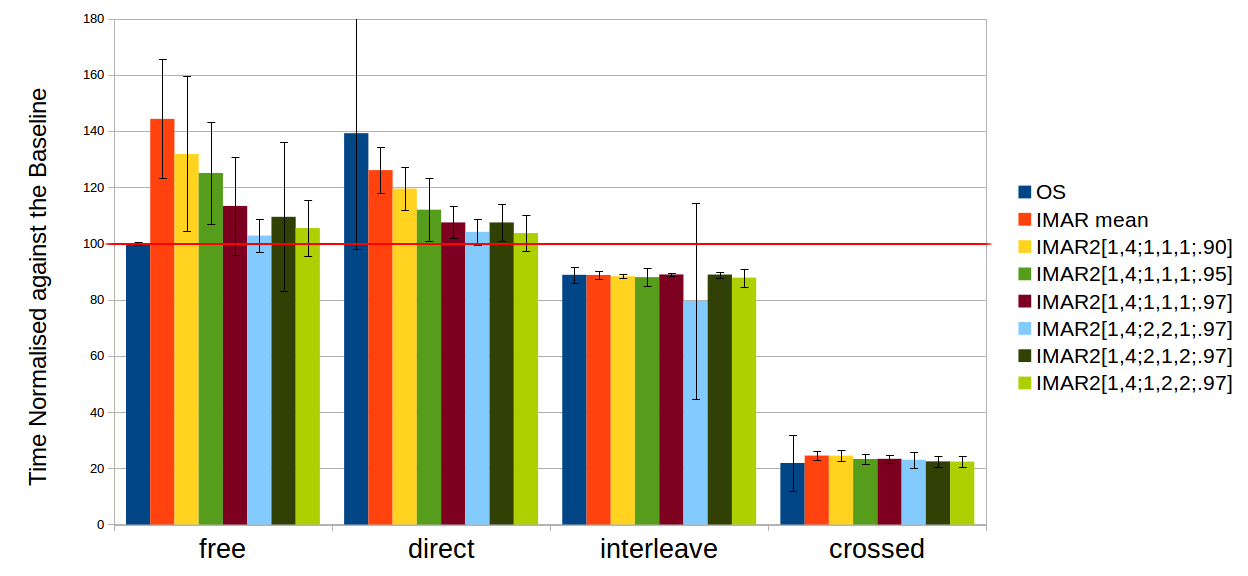}
 \end{center}
 \caption{Normalised execution times for the fastest lu.C instance in the \textbf{4 lu.C} test with IMAR\textsuperscript{2}.}
 \label{fig:4luC_1_a3}
 \end{figure*}

\begin{figure*}[tbp]
 \begin{center}
     \includegraphics[width=\textwidth]{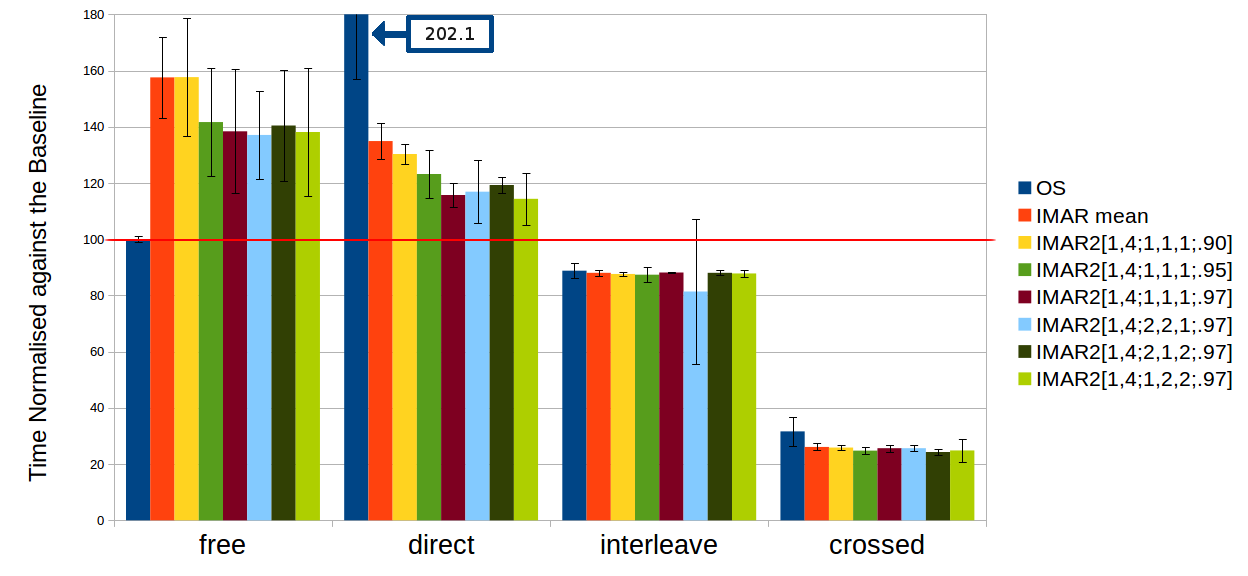}
 \end{center}
 \caption{Normalised execution times for the slowest lu.C instance in the \textbf{4 lu.C} test with IMAR\textsuperscript{2}.}
 \label{fig:4luC_4_a3}
 \end{figure*}

\begin{figure*}[tbp]
 \begin{center}
     \includegraphics[width=\textwidth,height=6.5cm]{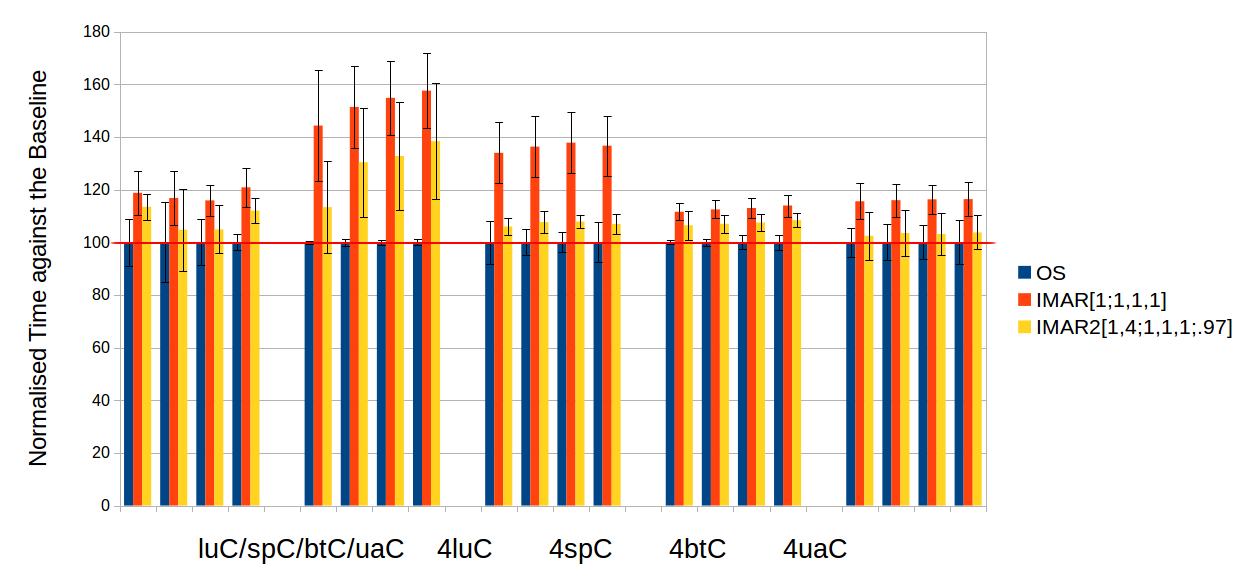}
 \end{center}
 \caption{Normalised execution times for \textsc{free} configuration with 4 nodes.}
 \label{fig:4n_mean_free}
 \end{figure*}

\begin{figure*}[tbp]
 \begin{center}
     \includegraphics[width=\textwidth,height=6.6cm]{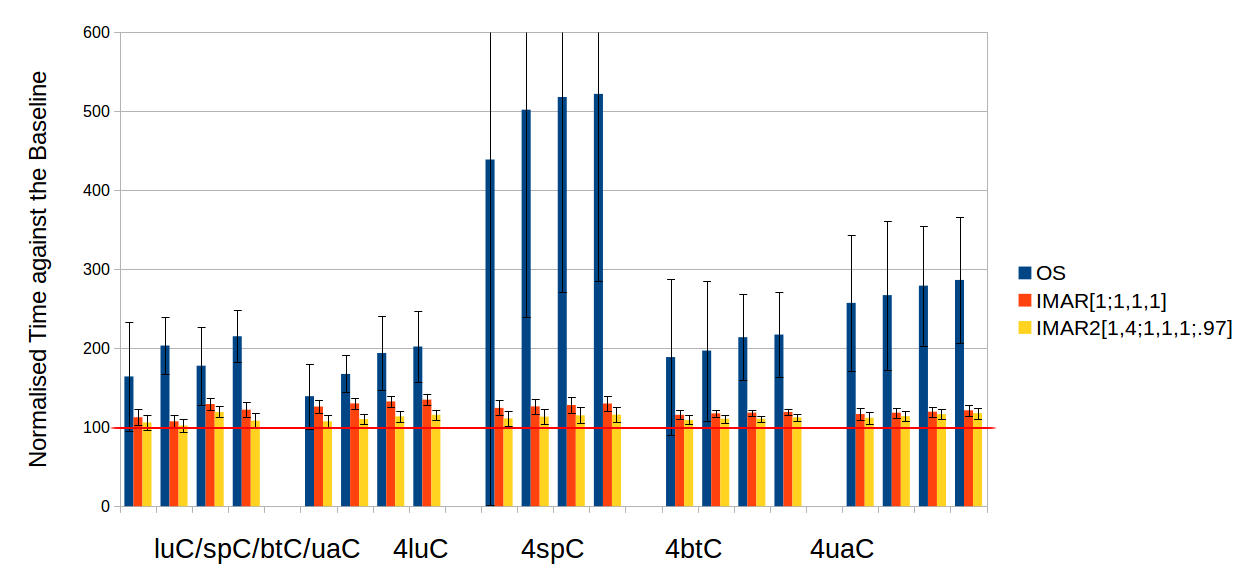}
 \end{center}
 \caption{Normalised execution times for \textsc{direct} configuration with 4 nodes.}
 \label{fig:4n_mean_direct}
 \end{figure*}

\begin{figure*}[tbp]
 \begin{center}
     \includegraphics[width=\textwidth,height=6.6cm]{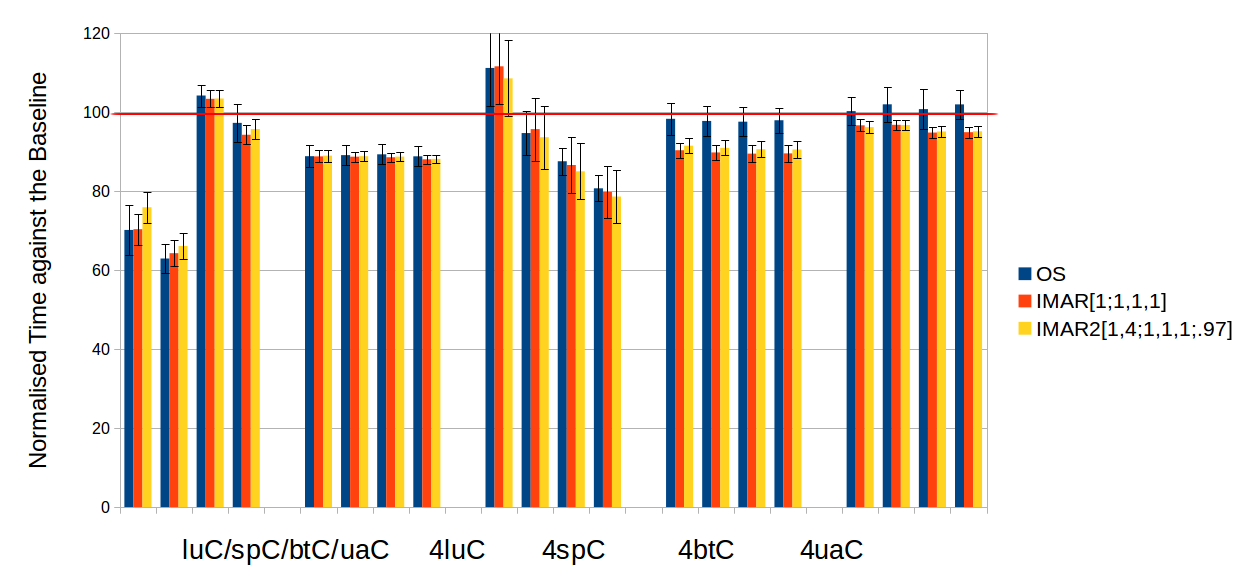}
 \end{center}
 \caption{Normalised execution times for \textsc{interleave} configuration with 4 nodes.}
 \label{fig:4n_mean_interleave}
 \end{figure*}

\begin{figure*}[tbp]
 \begin{center}
     \includegraphics[width=\textwidth,height=6.6cm]{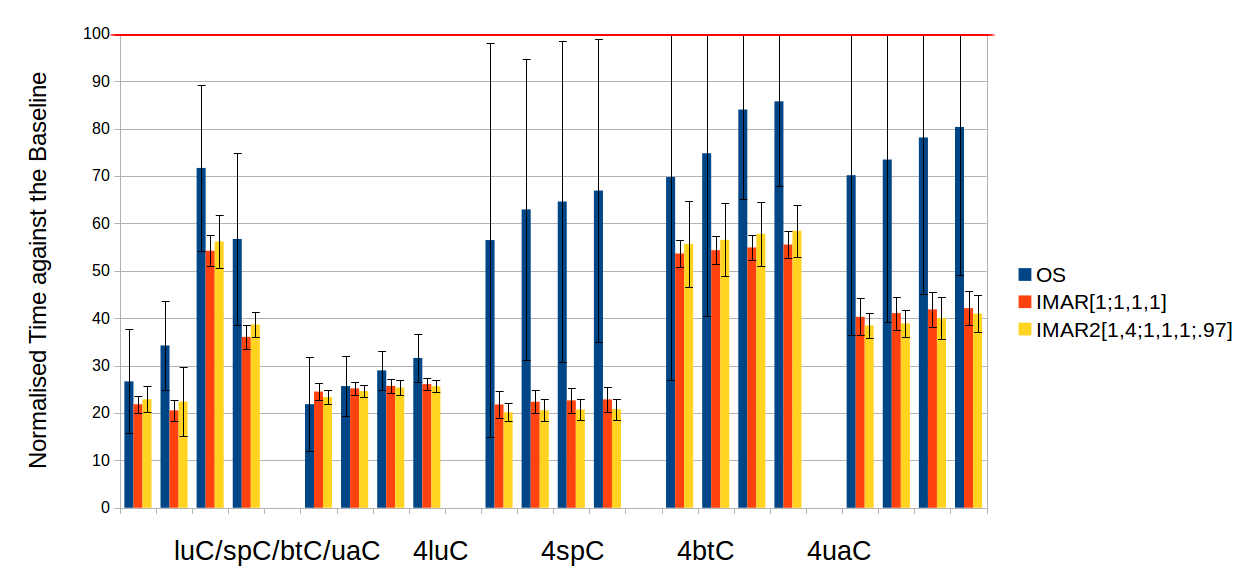}
 \end{center}
 \caption{Normalised execution times for \textsc{crossed} configuration with 4 nodes.}
 \label{fig:4n_mean_crossed}
 \end{figure*}

To compare IMAR\textsuperscript{2} with IMAR, the minimum and maximum times for IMAR\textsuperscript{2} were set to $T_{min}=1$ and $T_{max}=4$, so migrations would take place at approximately the same times as in the IMAR study.
In general, IMAR\textsuperscript{2} is superior to IMAR. For example, for combination \textbf{4 lu.C} (Figs.~\ref{fig:4luC_1_a3}~and~\ref{fig:4luC_4_a3}), as $\omega$ increases from $0.90$ to $0.97$, the loss of performance in \textsc{free} and \textsc{direct} tests is reduced, while in the \textsc{interleave} and \textsc{crossed} cases IMAR\textsuperscript{2} remains similar to the IMAR algorithm.
Figures~\ref{fig:4n_mean_free}--\ref{fig:4n_mean_crossed} show a closer look at the tests with only OS migration, compared to IMAR[1;1,1,1], and IMAR\textsuperscript{2}[1,4,2;1,1,1;0.97].
These are similar to previous figures, but the data were collated in a different way.
Results are shown for each benchmark instance, for every combination, for one given test.
With $\omega=0.97$, most cases show less than a 10\% loss of performance from the baseline \textsc{free} (Fig.~\ref{fig:4n_mean_free}) and \textsc{direct} (Fig.~\ref{fig:4n_mean_direct}) tests, and the performance increase from baseline \textsc{interleave} (Fig.~\ref{fig:4n_mean_interleave}) and \textsc{crossed} (Fig.~\ref{fig:4n_mean_crossed}) tests are similar or superior to the IMAR case.

 \section{Conclusions}
\label{sec:Conclusions}
Modern multicore systems present complex memory hierarchies, and make load balancing, data locality and thread affinity important issues to obtain high performance.
In this paper, thread migration algorithms, based on optimisation of 3DyRM parameters, were used to increase performance.
The proposed techniques improve execution times when thread locality is poor and the OS cannot improve thread placement during runtime.  
A multiobjective optimisation method, weighted product, is proposed to combine the 3DyRM parameters.

Using hardware counters, the performance of each thread in the system could be obtained in runtime with low overhead, and a tool was implemented to perform thread migration and allocation during runtime, applying different migration strategies and algorithms, tuned by a set of factors.

Two proposed migration algorithms were tested in a variety of scenarios.
The IMAR algorithm uses collected information about previous performance for each thread to guide thread migration decisions.
This algorithm was tested on a server using benchmarks from the NPB-OMP.
On complex systems, where NUMA effects are more pronounced, a poor allocation of threads and data can degrade performance by a factor of up to 5 or 6.
Given a poor distribution of threads and data, the OS by itself is not able to detect and correct it, which greatly influences performance.
The IMAR algorithm was able to improve execution 
By up to 70\%. However, small performance losses were obtained in cases where the thread configuration was initially good.

The IMAR\textsuperscript{2} algorithm can be considered a refining of the IMAR algorithm. It is based on the concept of evaluating the effects on the system total performance of previous migrations and acting accordingly.
Specifically, IMAR\textsuperscript{2} is based on IMAR, but adds rollback and changes in the period between migrations. This provides for greater tuning and performs better for those cases where migrations are unnecessary, while still improving the performance for cases with low initial performance. 

Generally, IMAR\textsuperscript{2} was superior to IMAR, which was superior to allowing the OS to self-optimise.

\section*{Acknowledgements}

This work was partially supported by the Ministry of Education and Science of Spain, FEDER funds under contract TIN2016-76373-P, and Xunta de Galicia, GRC 2014/018. It was developed in the framework of the European network HiPEAC, the Spanish network CAPAP-H6 and Galician networks R2016/045 and R2016/037.

 \bibliographystyle{elsarticle-num} 
 \bibliography{myRefs.bib}

\end{document}